\documentclass[12pt]{article}
\usepackage{amsfonts}
\usepackage{amsmath}
\usepackage{amssymb}
\usepackage{axodraw}
\usepackage{mathptmx}
\usepackage[dvips]{color}
\usepackage{epsfig}
\usepackage{graphicx}
\newcommand{\calA}{{\cal A}}
\newcommand{\calB}{{\cal B}}

\newcommand{\calL}{{\cal L}}

\newcommand{\calU}{{\cal U}}

\newcommand{\kvec}{{\bf k}}

\newcommand{\kslash}{/\!\!\!\!\!k}
\newcommand{\pslash}{/\!\!\!\!\!p}
\newcommand{\qslash}{/\!\!\!\!\!q}

\newcommand{\dslash}{/\!\!\!\!\!\partial}

\newcommand{\Br}{\textrm{Br}}

\newcommand{\GeV}{{\rm GeV}}
\newcommand{\TeV}{{\rm TeV}}
\newcommand{\cm}{{\rm cm}}

\newcommand{\ubar}{\bar{u}}

\begin{document}
\baselineskip=16pt

\pagenumbering{arabic}

\vspace{1.0cm}

\begin{center}
{\Large\sf Electromagnetic dipole moments and radiative decays of
particles from exchange of fermionic unparticles}
\\[10pt]
\vspace{.5 cm}

{Yi Liao\footnote{liaoy@nankai.edu.cn}}

{Department of Physics, Nankai University, Tianjin 300071}

\vspace{2.0ex}

{\bf Abstract}

\end{center}

We construct the propagator for a free fermionic unparticle field
from basic considerations of scale and Lorentz invariance. The
propagator is fixed up to a normalization factor which is required
to recover the result of a free massless fermion field in the
canonical limit of the scaling dimension. Two new features appear
compared to the bosonic case. The propagator contains both $\gamma$
and non-$\gamma$ terms, and there is a relative phase of $\pi/2$
between the two in the time-like regime for arbitrary scaling
dimension. This should result in additional interference effects on
top of the one known in the bosonic case. The non-$\gamma$ term can
mediate chirality flipped transitions that are not suppressed by a
light fermion mass but are enhanced by a large bosonic mass in
loops, compared to the pure particle case. We employ this last
feature to set stringent bounds on the Yukawa couplings between a
fermionic unparticle and an ordinary fermion through electromagnetic
dipole moments and radiative decays of light fermions.

\begin{flushleft}
PACS: 11.15.Tk, 14.80.-j, 13.40.Em, 13.35.-r

Keywords: unparticle, electromagnetic moment, radiative decay
\end{flushleft}

\newpage

Recently, Georgi suggested \cite{Georgi:2007ek} that a high energy
scale theory with a nontrivial infrared fixed point may manifest
itself at low energies as an effective degree of freedom that is
scale invariant. Since such an identity cannot be a particle of
definite mass, he termed it unparticle. The suggestion has
stimulated in the past year intense activities, exploring various
aspects of unparticles interacting with standard model particles.
Most studies, with very few exceptions
\cite{Luo:2007bq,Liaonote1,Henote1,Liaonote2,He:2008rv}, have dealt
with unparticles of integer spin. This is perhaps partly because
such unparticles can more easily couple as a standard model singlet
to particles and partly because their propagators as a necessary
ingredient for any physics analysis are known at the very start
\cite{Georgi:2007si,Cheung:2007ue,Cheung:2007ap}. On the other hand,
since unparticles must interact with particles to be physically
relevant, it should not be surprising that they could carry standard
model charges. There have been indeed interesting attempts in this
direction very recently \cite{Cacciapaglia:2007jq,Liao:2007fv,
Licht:2008ic,Galloway:2008jn,LewisLicht:2008mq}, though delicacies
arise with gauge interactions of unparticles due to the highly
nonlocal nature of the latter \cite{Liao:2007fv,Liaonote3}. In such
a circumstance, it is no more difficult to couple fermionic
unparticles to particles. Furthermore, as it is to be shown in this
brief note, the propagator for a fermionic unparticle can also be
constructed from basic considerations \cite{Liaonote1,Liaonote2}.
Though simple, the propagator brings in new features not seen in its
bosonic counterparts. It is also a purpose of this note to
investigate some of their phenomenological implications as
exemplified in chirality flipped processes.

The first attempt for a fermionic unparticle was made by Luo and Zhu
\cite{Luo:2007bq}. They treat its propagator as a Green's function
of effective operators and parameterize it in terms of unknown
constants through considerations of scale invariance and Lorentz
symmetry.
There are two points concerning that treatment. First, the
propagator of a massless fermion should be recovered as the
unparticle's scaling dimension goes to the canonical limit
\cite{Liaonote1,Luo:2007bq,Henote1}. Second, the unparticle field is
free when its interactions with particles are ignored at the leading
order in the low energy effective theory. Therefore, its propagator
should not contain unknown parameters just like its bosonic
counterparts except that the scaling dimension is fixed from a high
energy theory and that the absolute normalization is immaterial
\cite{Liaonote1,Liaonote2}. As we shall show below, the propagator
can indeed be constructed for a free fermionic unparticle field
through a similar procedure as for a canonical fermion field.

Similar to the canonical case, a free fermionic unparticle field as
a function in spacetime can be Fourier decomposed as
\cite{Liaonote1}:
\begin{eqnarray}
\calU(x)&=&\int\frac{d^4p}{(2\pi)^4}\Xi_d(p)\sum_s\left( a^s_p
u^s(p)e^{-ip\cdot x}+b^{s\dagger}_pv^s(p)e^{ip\cdot x}\right),\\
\Xi_d(p)&=&F(d)\theta(p^0)\theta(p^2) (p^2)^{d-\frac{5}{2}},
\end{eqnarray}
where the scaling property for an unparticle field of dimension $d$
has been taken into account. $F(d)$ is a normalization factor
appropriate for a fermionic $\calU$. The operators $a^s_p,~b^s_p$ of
dimension $(\frac{1}{2}-d)$ are assumed to satisfy the
anti-commutation relations, e.g.,
\begin{eqnarray}
\Xi_d(p)\Xi_d(q)\{a^r_p,a^{s\dagger}_q\}=\Xi_d(p)
(2\pi)^4\delta^4(p-q)\delta^{rs},
\end{eqnarray}
with the normalized single-unparticle states, e.g.,
\begin{eqnarray}
|p,s\rangle&=&\Xi_d(p)a^{s\dagger}_p|0\rangle,\\
\langle p,r|q,s\rangle&=& \Xi_d(p)(2\pi)^4\delta^4(p-q)\delta^{rs}.
\end{eqnarray}
$u^s(p),~v^s(p)$ are spinor wavefunctions of dimension $\frac{1}{2}$
to be built below.

The above decomposition would be completely similar to that for a
complex scalar unparticle field \cite{Liaonote1} (for a discussion
of the scalar case, see also \cite{Nikolic:2008ax}) in the absence
of $u^s(p),~v^s(p)$. The other factors each carry a dimension that
has the symmetry, $(d-\frac{1}{2})\leftrightarrow d$, with their
scalar counterparts. This suggests a natural choice for the
normalization factor, $F(d)=B(d-\frac{1}{2})$, where $B(d)$, usually
denoted as $A_d$, is the original normalization chosen by analogy in
Ref \cite{Georgi:2007ek}. This choice guarantees automatically that
the massless canonical propagator is recovered as $d\to\frac{3}{2}$.
We should emphasize that except for this welcome property,
normalization is arbitrary. While it affects apparent couplings, it
does not modify for example the cross section of a physical process.
This is obvious for unparticles appearing as an intermediate state
where the normalization convention cancels between couplings and
propagators, or in the final state where the cancelation occurs for
a squared single vertex and phase space. For unparticles appearing
in the initial state, the latter still holds true \cite{Henote2}:
their flux or density is convention dependent.

Although an unparticle has no dispersion relation, the standard
construction for a spinor wavefunction still works by the definition
of its Lorentz properties \cite{Liaonote2}. For a physical
unparticle with $p^2>0$, we go to its rest frame, where $\sqrt{p^2}$
plays the role of mass for a canonical fermion. A Lorentz boost then
gives the wavefunction for a general $p$ but with the same $p^2$ as
in the rest frame of course. The $u(p)$ thus constructed satisfies
the equations
\begin{eqnarray}
(\pslash-\sqrt{p^2})u(p)=0,~~~
\sum_{\textrm{spin}}u(p)\ubar(p)=\pslash+\sqrt{p^2},
\end{eqnarray}
and similarly for $v(p)$. Note that the Klein-Gordon equation
becomes a trivial identity due to the lack of dispersion relation.

We are now ready to compute the propagator from its definition:
\begin{eqnarray}
\tilde S_{F\alpha\beta}^\calU(x-y)=\langle
0|T\calU_\alpha(x)\overline\calU_\beta(y)|0\rangle,
\end{eqnarray}
where, using the relations elaborated so far,
\begin{eqnarray}
\langle 0|\calU_\alpha(x)\overline\calU_\beta(y)|0\rangle
&=&\int\frac{d^4p}{(2\pi)^4}\Xi_d(p)e^{-ip\cdot(x-y)}
\left(\pslash+\sqrt{p^2}\right)_{\alpha\beta}\nonumber\\
&=&+i(\dslash^x)_{\alpha\beta}\int\frac{d^4p}{(2\pi)^4}\Xi_d(p)
e^{-ip\cdot(x-y)}\nonumber\\
&&+\delta_{\alpha\beta}\int\frac{d^4p}{(2\pi)^4}\Xi_d(p)
(p^2)^{\frac{1}{2}}e^{-ip\cdot(x-y)}.
\end{eqnarray}
Similarly,
\begin{eqnarray}
\langle 0|\overline\calU_\beta(y)\calU_\alpha(x)|0\rangle
&=&-i(\dslash^x)_{\alpha\beta}\int\frac{d^4p}{(2\pi)^4}\Xi_d(p)
e^{ip\cdot(x-y)}\nonumber\\
&&-\delta_{\alpha\beta}\int\frac{d^4p}{(2\pi)^4}\Xi_d(p)
(p^2)^{\frac{1}{2}}e^{ip\cdot(x-y)}.
\end{eqnarray}

The non-$\gamma$ term of the propagator is worked out in detail in
Appendix to be,
\begin{eqnarray}
\delta_{\alpha\beta}\frac{F(d)}{2\sin(d\pi)}\int\frac{d^4p}{(2\pi)^4}
e^{-ip\cdot(x-y)}\frac{i}{(-p^2-i\epsilon)^{2-d}},\label{eq_non_gamma}
\end{eqnarray}
which is the same as the propagator for a scalar unparticle except
for the factors $F(d)$ vs $B(d)$. The $\gamma$ term is
\begin{eqnarray}
i(\dslash^x)_{\alpha\beta}\int\frac{d^4p}{(2\pi)^4}\Xi_d(p)
\left[\theta(x_0-y_0)e^{-ip\cdot(x-y)}+
\theta(y_0-x_0)e^{ip\cdot(x-y)}\right],
\end{eqnarray}
where the terms resulting from moving the derivatives across the
step functions sum to zero. The integral is again of the same form
as for a scalar field, which can be obtained from our previous
result by shifting $d\to d-\frac{1}{2}$ in all factors except
$F(d)$. Acting the derivative inside yields the term,
\begin{eqnarray}
\frac{F(d)}{2\sin[(d-\frac{1}{2})\pi]}
\int\frac{d^4p}{(2\pi)^4}e^{-ip\cdot(x-y)}
\frac{i\pslash}{(-p^2-i\epsilon)^{\frac{5}{2}-d}}.
\end{eqnarray}

To summarize, the propagator is \cite{Liaonote2},
\begin{eqnarray}
\tilde S_F^\calU(x-y)&=&\int\frac{d^4p}{(2\pi)^4}e^{-ip\cdot(x-y)}
S_F^\calU(p),\\
S_F^\calU(p)&=&\frac{iF(d)}{2\sin(d\pi)}\left[
\frac{1}{(-p^2-i\epsilon)^{2-d}}-\tan(d\pi)
\frac{\pslash}{(-p^2-i\epsilon)^{\frac{5}{2}-d}}\right].
\end{eqnarray}
Using the convention,
\begin{eqnarray}
F(d)=B(d-1/2)=\frac{16\pi^{\frac{5}{2}}}{(2\pi)^{2d-1}}
\frac{\Gamma(d)}{\Gamma(d-\frac{3}{2})\Gamma(2d-1)},
\end{eqnarray}
it can be checked that $S_F^\calU$ recovers the standard result for
a massless fermion as $d\to\frac{3}{2}$. The discontinuity in
$S_F^\calU(p)$ across the cut for $p^2>0$ is found to be, using
$(-|r|\pm i\epsilon)^\alpha=|r|^\alpha e^{\mp i\pi\alpha}$,
\begin{eqnarray}
F(d)(p^2)^{d-\frac{5}{2}}(\pslash+\sqrt{p^2}),
\end{eqnarray}
exactly as expected.

Two new features are worth mention. It is well known that the
bosonic unparticle propagator develops an imaginary part in the
time-like regime for non-integral dimension $d$. This has been shown
to cause interesting interference between unparticle and particle
contributions to a physical process \cite{Georgi:2007si}. In
addition to this, we find that in the fermionic case the two terms
in the propagator has a relative phase of $\pi/2$ in the time-like
regime for any $d$. This can result in even more interesting
phenomena. Furthermore, the mass term in an ordinary fermionic
propagator is now replaced by a momentum dependent one free of the
$\gamma$ matrices. It is known in the case of particles that the
mass of an intermediate fermion can flip the chirality of its
connected initial or final fermions. When the latter are light, the
contribution of such a term is naturally suppressed either directly
by a light intermediate fermion, or indirectly by the small mixing
between light and heavy fermions in the opposite case. In the
present case where a fermionic unparticle appears in the
intermediate state, this non-$\gamma$ term is generally not
suppressed since its momentum can be typically of the order of a
heavy particle mass involved in the intermediate state. In the
second part of this work, we explore its implications on chirality
flipped processes of light fermions, namely the electromagnetic
transitions induced by interactions with fermionic unparticles.

Consider the effective interaction,
\begin{eqnarray}
\calL_{\rm int}^\calU&=&\Lambda_\calU^{\frac{3}{2}-d}
\overline{\calU}(a_j+b_j\gamma_5)\psi_j\varphi
+\Lambda_\calU^{\frac{3}{2}-d}\bar\psi_j
(a_j^*-b_j^*\gamma_5)\calU\varphi^\dagger,\label{eq_L_U}
\end{eqnarray}
where $\varphi$, $\psi_j$ are the ordinary scalar and fermion fields
with mass $m$ and $m_j$ respectively. $\Lambda_\calU$ characterizes
the energy scale of unparticle physics, and $a_j,~b_j$ are unknown
pure numbers. We have deliberately chosen $\calU$ to be electrically
neutral to avoid its gauge interactions as the effect on chirality
flip can be easily seen via the above interaction together with
ordinary QED. Then, in units of $e>0$, the electric charges are,
$Q(\varphi)=-Q(\psi_j)\equiv-Q_j$.

\begin{center}
\begin{picture}(200,90)(0,0)
\SetOffset(50,30)%
\ArrowLine(0,0)(20,0)
\Line(20,0.8)(80,0.8)\Line(20,-0.8)(80,-0.8)\ArrowLine(48,0)(52,0)
\ArrowLine(80,0)(100,0)%
\DashArrowArc(50,0)(30,0,90){3}\DashArrowArc(50,0)(30,90,180){3}%
\Photon(50,30)(50,50){-3}{3}%
\Text(50,58)[]{$\mu$}\Text(10,-8)[]{{\footnotesize$p,1$}}
\Text(90,-8)[]{{\footnotesize$p\!-\!q,2$}}%
\Text(36,15)[]{{\footnotesize$k$}}
\Text(50,8)[]{{\footnotesize$\calU$}}%
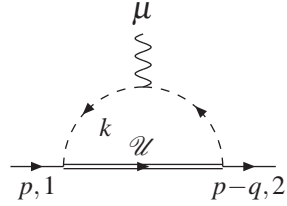
\Text(50,-25)[]{Figure 1. One-loop diagram for
$\psi_1\to\psi_2\gamma$.}
\end{picture}
\end{center}

The interaction induces the flavor-changing, chirality-flipped
transition, $\psi_1\to\psi_2\gamma$, via the graph shown in Fig. 1,
where the arrowed double line indices the $\calU$ field. The
amplitude is,
\begin{eqnarray}
i\calA_\mu&=&\frac{Q_1e\Lambda_\calU^{3-2d}F(d)}{2\sin(d\pi)}
\int\frac{d^4k}{(2\pi)^4}
\frac{(2k+q)_\mu(a_2^*-b_2^*\gamma_5)}{[k^2-m^2][(k+q)^2-m^2]}
\nonumber\\
&\times&\left[\frac{1}{[-(k+p)^2-i\epsilon]^{2-d}}
-\tan(d\pi)\frac{\kslash+\pslash}{[-(k+p)^2-i\epsilon]^{\frac{5}{2}-d}}
\right](a_1+b_1\gamma_5).
\end{eqnarray}
Keeping only terms that contribute to the on-shell amplitude yields
in the limit $m\gg m_{1,2}$,
\begin{eqnarray}
\calA_\mu&=&-\frac{1}{(4\pi)^2}
\frac{Q_1e\Lambda_\calU^{3-2d}F(d)}{2\sin(d\pi)}m^{2(d-2)}
\calB_\mu,\\
\calB_\mu&=&\tan(d\pi)\Gamma(d+1/2)\Gamma(7/2-d)
\frac{p_\mu(\qslash-2\pslash)}{6m}\nonumber\\
&&\times[(a_1a_2^*+b_1b_2^*)+(a_2^*b_1+a_1b_2^*)\gamma_5]\nonumber\\
&+&\Gamma(d)\Gamma(3-d)p_\mu[(a_1a_2^*-b_1b_2^*)+
(a_2^*b_1-a_1b_2^*)\gamma_5].
\end{eqnarray}
When sandwiched between the external spinors, it becomes the
standard dipole form,
\begin{eqnarray}
\calB_\mu&=&\tan(d\pi)\frac{1}{6m}\Gamma(d+1/2)\Gamma(7/2-d)\nonumber\\
&\times&\left[-\bar m(a_1a_2^*+b_1b_2^*)i\sigma_{\mu\nu}q^\nu
+\Delta m(a_2^*b_1+a_1b_2^*)i\sigma_{\mu\nu}q^\nu\gamma_5\right]
\nonumber\\
&+&\Gamma(d)\Gamma(3-d)\frac{1}{2}\left[
(a_1a_2^*-b_1b_2^*)i\sigma_{\mu\nu}q^\nu+ (a_2^*b_1-a_1b_2^*)
i\sigma_{\mu\nu}q^\nu\gamma_5\right],\label{eq_B}
\end{eqnarray}
where $\bar m=(m_1+m_2)/2,~\Delta m=(m_1-m_2)/2$.

For $\psi_1=\psi_2$, the above amplitude gives directly the
electromagnetic dipole moments. The anomalous magnetic moment
(coefficient of $-ei\sigma_{\mu\nu}q^\nu/(2m_1)$ in $\calA_\mu$) is,
\begin{eqnarray}
a_{\psi_1}&=&\frac{1}{2}(g_{\psi_1}-2)\nonumber\\
&=&Q_1\frac{2m_1}{(4\pi)^2}\frac{\Lambda_\calU^{3-2d}F(d)}{2\sin(d\pi)}m^{2(d-2)}
\left[\frac{1}{2}\Gamma(d)\Gamma(3-d)(|a_1|^2-|b_1|^2)\right.\nonumber\\
&&\left. -\tan(d\pi)\frac{m_1}{6m}\Gamma\left(d+\frac{1}{2}\right)
\Gamma\left(\frac{7}{2}-d\right)(|a_1|^2+|b_1|^2)\right],\label{eq_a}
\end{eqnarray}
while the coefficient of $-\sigma_{\mu\nu}q^\nu\gamma_5$
(corresponding to $\calL^{\rm EDM}=-\frac{1}{2}d\bar\psi
i\sigma_{\mu\nu}\gamma_5\psi F^{\mu\nu}$) gives the electric dipole
moment,
\begin{eqnarray}
d_{\psi_1}=-Q_1\frac{1}{(4\pi)^2}
\frac{e\Lambda_\calU^{3-2d}F(d)}{2\sin(d\pi)}m^{2(d-2)}
\Gamma(d)\Gamma(3-d)\Im(a_1^*b_1).\label{eq_d}
\end{eqnarray}

It is evident from the above results that the non-$\gamma$ term in
the unparticle propagator enhances the amplitude by a factor of
$m/m_j$ which is large for transitions of light fermions. This
occurs because the chirality flip implemented usually by a mass in
the pure particle case is now operated by a momentum, which in loops
can be of order the mass scale of heavy particles; in short, an
internal fermion mass is effectively traded for a large boson mass
for chirality flip. For better appreciation of this effect, we
record here the dipole moments of $\psi$,
\begin{eqnarray}
a_\psi&=&-\frac{Q_\psi}{(4\pi)^2}\frac{m_\psi}{m^2}\left[
m_\chi(|a|^2-|b|^2)+\frac{1}{3}m_\psi(|a|^2+|b|^2)\right],
\nonumber\\
d_\psi&=&\frac{Q_\psi e}{(4\pi)^2}\frac{m_\chi}{m^2}\Im(a^*b),
\end{eqnarray}
due to pure particle interactions,
\begin{eqnarray}
\calL_{\rm int}^\chi=\bar\chi(a+b\gamma_5)\psi\varphi +\bar
\psi(a^*-b^*\gamma_5)\chi\varphi^\dagger,
\end{eqnarray}
with $Q(\varphi)=-Q(\psi)\equiv -Q_\psi,~Q(\chi)=0$. It is assumed
again that $m\gg m_{\psi,\chi}$. Note that the $m_\psi$ suppressed
term in $a_\psi$ can be obtained from the corresponding term in the
unparticle result in the limit $d\to\frac{3}{2}$, while the $m_\chi$
suppressed terms in $a_\psi$ and $d_\psi$ are replaced in the
unparticle case by the enhanced ones.

We can obtain some stringent constraints on the couplings appearing
in (\ref{eq_L_U}). Since the non-$\gamma$ term overwhelmingly
dominates, we only retain its contribution in
(\ref{eq_B},\ref{eq_a},\ref{eq_d}) for our numerical analysis. We
start with the dipole moments:
\begin{eqnarray}
a_{\psi_1}&=&Q_1f(d)\left(\frac{m}{\Lambda_\calU}\right)^{2d-3}
\frac{m_1}{m}(|a_1|^2-|b_1|^2),\\
\frac{d_{\psi_1}}{e}&=&-Q_1f(d)\left(\frac{m}{\Lambda_\calU}\right)^{2d-3}
\frac{1}{m}\Im(a_1^*b_1),
\end{eqnarray}
where
\begin{eqnarray}
f(d)=\frac{1}{2^{2d}\pi^{2d-\frac{3}{2}}\sin(d\pi)}
\frac{[\Gamma(d)]^2\Gamma(3-d)}{\Gamma(d-\frac{3}{2})\Gamma(2d-1)}.
\end{eqnarray}
The current potential discrepancies between experiments and standard
model expectations for leptons' magnetic moments are, $\delta
a_i=a_i^{\rm expt}-a_i^{\rm SM}$,
\begin{eqnarray}
|\delta a_e|<15\times 10^{-12}~\cite{odom},~~~ \delta
a_\mu=22(10)\times 10^{-10}.
\end{eqnarray}
Here, unless otherwise stated, we use the numbers of the Particle
Data Group, version 2006. We assume the gap is filled by the
unparticle contribution. The most precise experimental results for
the electric dipole moments are those of the electron, muon and
neutron:
\begin{eqnarray}
d_e&=&(0.07\pm 0.07)\times 10^{-26}~{\rm e~}\cm,\nonumber\\
d_\mu&=&(3.7\pm 3.4)\times 10^{-19}~{\rm e~}\cm,\\
d_n&<&0.63\times 10^{-25}~{\rm e~}\cm.\nonumber
\end{eqnarray}
Since the standard model contributions to the above are ignorable,
we assume that the unparticle saturates the central values or the
upper bound. For the purpose of illustration, we take
$\Lambda_\calU=1~\TeV,~m=200~\GeV$, and $d\in(1.5,2.0)$. The
constraints on the couplings are shown in table 1. We have assumed
that the neutron electric dipole is dominated by those of the $u,~d$
quarks, and ignore factors of $2/3,~1/3$ in forming the neutron's
moment from those of quarks. It is evident that the constraints are
rather stringent for the most precisely measured quantities, i.e.,
$d_e,~d_n$. We remind that in the case of pure particle interactions
the Yukawa couplings' contribution to dipole moments is completely
ignorable.
\begin{table}
\begin{center}
\begin{tabular}{|r|l|l|l|l|}
\hline
$d$                &$1.6$              &$1.7$ &$1.8$ &$1.9$\\
\hline
$10^3(|a|^2-|b|^2)_e$  &$7.3$&$6.5$&$6.5$&$5.4$\\
$10^3(|a|^2-|b|^2)_\mu$&$5.2$&$4.6$&$4.6$&$3.8$\\
$-10^9        \Im(a^*b)_e$&$8.9$&$7.8$&$7.9$&$6.5$\\
$      -\Im(a^*b)_\mu$&$4.7$&$4.1$&$4.2$&$3.4$\\
$-10^7        \Im(a^*b)_n$&$8.0$&$7.0$&$7.1$&$5.9$\\%
\hline
\end{tabular}
\caption{Couplings needed to saturate the deviations in $a_j$ or the
measured results for $d_j$.}\label{tab_1}
\end{center}
\end{table}

Now we move to the flavor changing electromagnetic transitions. The
dominant term for $\psi_1\to\psi_2\gamma$ is,
\begin{eqnarray}
\calA_\mu&=&-\frac{f(d)Q_1e}{2m}\left(\frac{m}{\Lambda_\calU}\right)^{2d-3}
\left[(a_1a_2^*-b_1b_2^*)i\sigma_{\mu\nu}q^\nu+ (a_2^*b_1-a_1b_2^*)
i\sigma_{\mu\nu}q^\nu\gamma_5\right],
\end{eqnarray}
yielding the rate,
\begin{eqnarray}
\Gamma(\psi_1\to\psi_2\gamma)=2^{-7}m_1\alpha
\left[f(d)Q_1\left(\frac{m}{\Lambda_\calU}\right)^{2d-3}\frac{m_1}{m}\right]^2
X_{12},
\end{eqnarray}
where $m_1\gg m_2$ is assumed in kinematics, and
\begin{eqnarray}
X_{ij}=|a_ia_j^*-b_ib_j^*|^2+|a_j^*b_i-a_ib_j^*|^2.
\end{eqnarray}
The most stringent experimental bounds are those on lepton flavor
changing radiative decays:
\begin{eqnarray}
&&\Br(\mu\to e\gamma)<1.2\times 10^{-11}~\cite{mega},\nonumber\\
&&\Br(\tau\to\mu\gamma)<4.5\times 10^{-8}~\cite{belle},~\Br(\tau\to
e\gamma)<1.2\times 10^{-7}~\cite{belle}.
\end{eqnarray}
Again we assume that these transitions are dominated by the
fermionic unparticles. This produces the bounds on various $X_{ij}$
shown in table 2. Note that
$\Br(\tau\to\nu_\tau\ell\bar\nu_\ell)\sim 17\%$ ($\ell=e,~\mu$) has
been taken into account. The bound on $X_{\mu e}$ is enhanced not
only by a more precise measurement but also by a less power
dependence of the fermion mass in the transition amplitude compared
to the particle case. The bound on $X_{bs}$ from the decay $b\to
s\gamma$ is less stringent.
\begin{table}
\begin{center}
\begin{tabular}{|r|l|l|l|l|}
\hline
$d$               &$1.6$&$1.7$&$1.8$ &$1.9$\\
\hline
$10^{12}X_{\mu e}$&$3.3$&$2.6$&$2.7$&$1.8$\\
$ 10^5X_{\tau e}$ &$5.3$&$4.1$&$4.2$&$2.9$\\
$ 10^5X_{\tau\mu}$&$2.0$&$1.5$&$1.6$&$1.1$\\
\hline
\end{tabular}
\caption{Bounds on the combination of couplings $X_{ij}$. Same input
parameters as in table \ref{tab_1}.}
\end{center}
\end{table}

We have worked out the propagator of a free fermionic unparticle
from basic considerations of scale invariance and Lorentz symmetry.
It has a correct particle limit as its scaling dimension goes to the
canonical limit. The propagator has a $\gamma$ dependent term as
naively expected, and a momentum-dependent non-$\gamma$ term that
would correspond to the mass term in the particle case. There is a
{\it relative} phase between the two terms in the time-like regime,
which can result in interesting interference phenomenon on top of
the one known already for a bosonic unparticle. We pointed out that
the non-$\gamma$ term can cause chirality flip that is not
suppressed by a fermion mass, in contrast to the case of particles.
Instead, when appearing in loops, the term is traded for a large
mass of virtual bosonic particles, and can thus be dangerous for
precisely measured chirality-flipped quantities, like
electromagnetic dipole moments and radiative decays of light
fermions. We have employed this to set stringent bounds on the
mixing Yukawa couplings between fermionic unparticles and light
particles. For the diagonal combinations of couplings, the anomalous
magnetic moments of the electron and muon give similar constraints
on their magnitude, while the electron's electric dipole moment
yields the best constraint on the imaginary part. For the
non-diagonal combinations, the radiative decay of the muon sets a
limit that is several orders of magnitude stronger than the tau
decays.

\vspace{0.5cm}
\noindent %
{\bf Acknowledgement} I thank Xiao-Gang He for participating the
work at its early stage and for many helpful discussions and
encouragements. This work is supported in part by the grants
NCET-06-0211 and NSFC-10775074.

\vspace{0.5cm}
\noindent %
{\it Note added.} As our phenomenological analysis was going on, we
received a preprint \cite{basu} in which the propagator of a
fermionic unparticle was worked out with the help of a spectral
function \cite{Georgi:2007si}. The derivation is different but the
result is the same as obtained here, up to a different convention
for the normalization that also has the correct canonical limit. The
main concern of that work is with gauge interactions of unparticles
and their contribution to the $\beta$ function, following the work
in Refs. \cite{Cacciapaglia:2007jq,Liao:2007fv}.

\vspace{0.5cm}
\noindent %
{\bf Appendix} Derivation of equation (\ref{eq_non_gamma})

We derive the equation for completeness. Using
\begin{eqnarray}
\theta(\eta)=\frac{i}{2\pi}\int_{-\infty}^{\infty}
\frac{ds}{s+i\epsilon}e^{-is\eta},
\end{eqnarray}
and making appropriate changes of parameters, the first term becomes
\begin{eqnarray}
\delta_{\alpha\beta}F(d)\frac{i}{2\pi}\int\frac{d^4k}{(2\pi)^4}
\int_{|\kvec|}^{\infty}\frac{dt}{(k_0+i\epsilon)-t}
\left(t^2-\kvec^2\right)^{d-2} e^{-ik\cdot(x-y)}.
\end{eqnarray}
The second term only differs in the sign of the exponent. They sum
to
\begin{eqnarray}
-\delta_{\alpha\beta}F(d)
\frac{i}{2\pi}\int\frac{d^4k}{(2\pi)^4}e^{-ik\cdot(x-y)}
\int_0^{\infty}d\rho\frac{\rho^{d-2}}{\rho-k^2-i\epsilon}.
\end{eqnarray}
Using Cauchy theorem, the $\rho$ integral can be evaluated along the
ray starting from the origin and passing the point
$\rho=-\rho_0=-(k^2+i\epsilon)$. The term becomes then
\begin{eqnarray}
-\delta_{\alpha\beta}F(d)\frac{1}{2\pi}\Gamma(2-d)\Gamma(d-1)
\int\frac{d^4k}{(2\pi)^4}e^{-ik\cdot(x-y)}\frac{i}{(-k^2-i\epsilon)^{2-d}},
\end{eqnarray}
which is (\ref{eq_non_gamma}) upon using the relation
\begin{eqnarray}
-\frac{1}{2\pi}\Gamma(2-d)\Gamma(d-1)=\frac{1}{2\sin(d\pi)}.
\end{eqnarray}


\end{document}